\documentclass[runningheads]{llncs}

\usepackage{graphicx}
\usepackage{xcolor}
\usepackage{enumitem,kantlipsum}
\usepackage{hyperref}
\usepackage{url}
\usepackage{algorithm,algorithmic}
\usepackage{amsmath,amssymb,amsfonts}
\usepackage{multicol}
\usepackage{multirow}
\usepackage{adjustbox}

\begin{document}

\title{Ontology-Compliant Knowledge Graphs}

\author{Zhangcheng Qiang\orcidID{0000-0001-5977-6506}}

\authorrunning{Z. Qiang}

\institute{School of Computing, Australian National University, 108 North Road, Acton, Canberra, ACT 2601, Australia \\ \email{qzc438@gmail.com}}

\begingroup
\renewcommand{\thefootnote}{}
\footnotetext[0]{Category: Early Stage PhD.}
\footnotetext[1]{\textcopyright\ The Author(s), under exclusive license to Springer Nature Switzerland AG 2023.}
\endgroup

\maketitle

\begin{abstract}

Ontologies can act as a schema for constructing knowledge graphs (KGs), offering explainability, interoperability, and reusability. We explore \emph{ontology-compliant} KGs, aiming to build both internal and external ontology compliance. We discuss key tasks in ontology compliance and introduce our novel term-matching algorithms. We also propose a \emph{pattern-based compliance} approach and novel compliance metrics. The building sector is a case study to test the validity of ontology-compliant KGs. We recommend using ontology-compliant KGs to pursue automatic matching, alignment, and harmonisation of heterogeneous KGs.


\end{abstract}

\section{Introduction and Motivation}

An ontology is typically used as the backbone for constructing a KG, building so-called \emph{ontology-based KGs}. In this setting, the ontology and the KG are often treated as independent functional components. An ontology provides a knowledge-oriented graph schema (i.e., TBox), whereas a KG represents the corresponding data-driven instances (i.e., ABox). With the proliferation of KGs in real-world applications, problems arise when data in the KG is generated for different user requirements. The ontology is likely to be incompatible with the data in the KG because ABox assertions may extend or be incomplete with respect to the ontology. While ABox contents can be adapted to suit a TBox, for interoperability amongst independent TBoxes, an ABox that is compatible with a number of TBoxes may be needed. Such overarching TBoxes should support conversion and exchange for cross-KG harvesting and federated searches.

\begin{figure}[htbp]
\centerline{\includegraphics[width=1\linewidth]{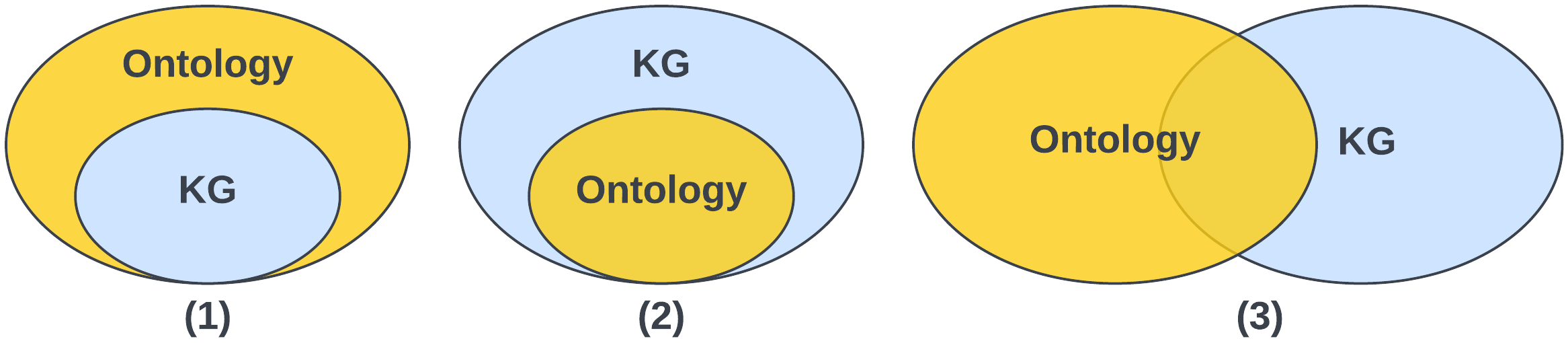}}
\caption{Non-compliance between KG and its ontology.}
\label{fig: ontology and KG}
\end{figure}

Fig.~\ref{fig: ontology and KG} illustrates three types of non-compliance between KG and its ontology. (1) The ABox in the KG only covers a small amount of TBox terminologies, and its ontology has many unused classes and properties. (2) The ABox in the KG contains more information than the TBox terminologies, and many terms in the KG cannot find appropriate classes and properties in its ontology. (3) In a combination of (1) and (2), the ABox in the KG and the TBox in the ontology are mismatched and overlapped on both sides. Many application tasks, for example, KG embedding and ontology learning, are hampered by non-compliance between KG and its ontology. When using KG embedding for ontology-based KGs, unused classes and properties in the ontology are noisy data. This results in inaccurate embeddings for KG terms. Ontology learning is the task of using KG to infer ontology classes and properties. KG data is diverse in nature; thus, the new ontology classes and properties learnt from KG can be different according to KG instances. These task-specific ontology classes and properties may violate the FAIR (i.e., Findability, Accessibility, Interoperability, and Reusability) principle and can be challenging to map and integrate into the original ontology.

Current work mainly focuses on either schema matching (i.e., TBox-TBox compliance) or instance matching (i.e., ABox-ABox compliance). TBox-ABox compliance is underexplored. While it is questionable whether the TBox is always compliant with the ABox, ontology-based KGs assume they are compliant by nature (excluding or ignoring the three types of non-compliance). For this reason, there is rarely a compliance check in popular KG and ontology modelling libraries or editors (e.g., RDFLib~\cite{boettiger2018rdflib}, Protégé~\cite{musen2015protege}, and TopBraid Composer~\cite{topbraid2022}). Even within the Ontology Alignment Evaluation Initiative (OAEI)~\cite{oaei}, to the best of our knowledge, we cannot find tools available to track these mismatches and overlaps between ABox in the KG and TBox in its corresponding ontology.

\section{State of the Art}

\textbf{Ontology-Based KGs} describe the traditional design for using ontologies with KGs, whereby the ontology serves as the schema for the KGs. KGs are generated using the classes and properties pre-defined in the ontology. In this setting, ontology-based KGs assume the ontology has established well-defined concepts, taxonomies, relationships, and domain axioms. Compared with ontology-less KGs, ontology-based KGs provide more formal representations for data understanding, organisation, and integration. They also enable improved logical reasoning, empowered reuse, and enhanced interoperability between different downstream applications. However, a complete ontology is almost impossible. Ontology is built on the Open World Assumption (OWA). We cannot assume an ontology has captured all domain concepts because the absence of concepts is not non-existence (i.e., these concepts may exist in other ontologies). A ``well-defined'' ontology also requires solid verification and validation. There is no gold standard for dealing with individual differences among opposing viewpoints.

\noindent\textbf{Ontology-Aware KGs} follow a reverse way of using ontologies with KGs. Conceptual components learnt from KGs are used to build or evolve the original ontology. The paradigm of ontology-aware KGs assumes the KG data is noiseless. There are two directions for constructing ontology-aware KGs. (1) Ontology reshaping is applied to data in the KG only covers part of the concepts in the ontology. The goal of ontology reshaping is to create a data-oriented local schemata that preserves the domain ontology knowledge while removing unused nodes~\cite{zhou2022enhancing,zhou2022ontology}. (2) Ontology enrichment is used where data exists in the KG that is not covered by the ontology. In this case, the new concepts and relationships learnt from the KG are registered as new classes and properties in the ontology~\cite{hurlburt2021knowledge,zhao2019learning}. While ontology-aware KGs achieve partial compliance between the KG and its ontology, they still have some limitations. Concepts that have been locally reshaped and redefined are task-specific, with limited sharing and reusing capabilities. Moreover, ontology-aware KGs cannot track the poly-ontological representation of KGs as they only match one KG to its corresponding ontology.

\section{Problem Statement and Contributions}

\begin{figure}[htbp]
\centerline{\includegraphics[width=1\linewidth]{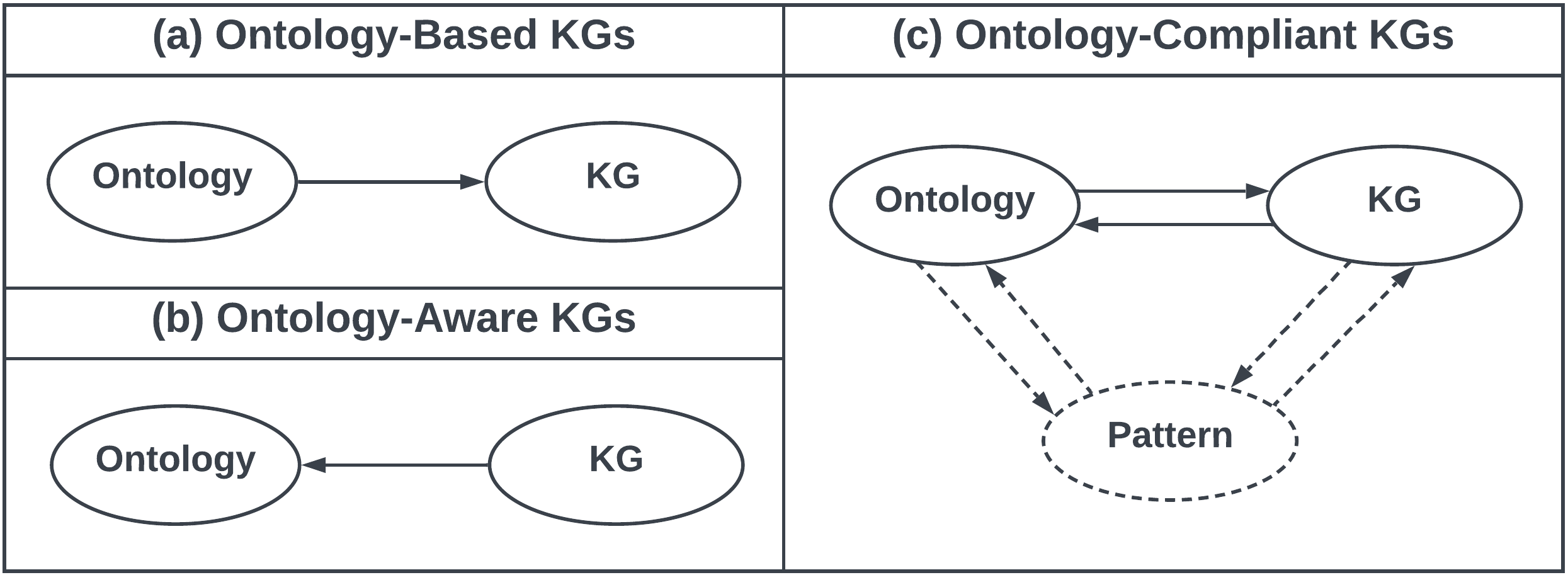}}
\caption{The difference between ontology-compliant KGs and the other two types.}
\label{fig: ontology-type-KGs}
\end{figure}

In the real world, KGs and ontologies are mostly incomplete. Neither ontology-based KGs nor ontology-aware KGs could fully handle the compliance issue between KGs and ontologies. We plan to propose \textbf{Ontology-Compliant KGs} to fill this gap. ``Compliant'' here has two aspects: (1) The terms used in KG are in line with the definition provided by the ontology. Mismatched terms in the KG are replaced with the most relevant classes and properties defined in the original ontology. (2) The size of the ontology complies with the information coverage of the KG. There are no unused classes or properties. In this work, we also extend this definition to be ontology compliant across KGs. Joint learning, vector embedding methods, and pattern-based engineering concepts are employed to achieve the goal of both internal and external compliance between KGs and ontologies. Fig.~\ref{fig: ontology-type-KGs} shows the difference between ontology-compliant KGs and the other two types of ontology-related KGs. While ontology-based and ontology-aware KGs only consider a one-way connection, in ontology-compliant KGs, the link between ontology and KG is bidirectional and can be bridged by their patterns (details are described in Section~\ref{sec: results}).

\noindent{\textbf{Hypothesis}} Ontology-compliant KGs have the following unique features:

\begin{enumerate}[wide, noitemsep, topsep=0pt, labelindent=0pt, label*=\textbf{H\arabic*}]
\item Given an ontology and a baseline KG, ontology-compliant KGs can eliminate the unused classes and properties in the ontology and reduce misdefined terms in the KG (interpreted as Ontology Compliance \emph{within} KG).
\item Given a set of ontologies and a baseline KG, ontology-compliant KGs allow automatic transmission from one schema to another (interpreted as Ontology Compliance \emph{over} KGs).
\item Given a set of ontologies and a baseline KG, ontology-compliant KGs allow different ontology fragment representations via a pattern-based approach. These ontology fragments are provided with multiple criteria for integration, evaluation, and selection (interpreted as \emph{Pattern-based} Compliance).
\end{enumerate}

\noindent\textbf{Research Questions} We formulate the related research questions:

\begin{enumerate}[wide, noitemsep, topsep=0pt, labelindent=0pt, label*=\textbf{RQ\arabic* (wrt H\arabic*)}]
\item How to reconstruct ontology-based KGs into ontology-compliant KGs, while retaining critical information and primary inference capability but eliminating unused nodes and reducing misdefined nodes?
\item How to enable schema-free KGs that can be compliant with multiple ontologies, using the ontology-compliant KGs to automate and optimise the ontology alignment and matching process?
\item How to select the most compliant set of ontology fragments for KGs? How to capture the different ontology fragment representations using a pattern-based approach, and evaluate them according to sound criteria from different useful perspectives?
\end{enumerate}

\section {Research Methodology and Approach}

Details of preliminary results based on the research methodology and approach are described in Section~\ref{sec: results}. This PhD aims to define a generalised approach to constructing ontology-compliant KGs. We propose to classify three stages of compliance in ontology-compliant KGs, namely (1) Ontology Compliance \emph{within} KG, (2) Ontology Compliance \emph{over} KGs, and (3) \emph{Pattern-Based} Compliance. In each stage, we intend to address the hypothesis and its related research question. The ``building domain'' is selected as a case study. We design, implement, and evaluate our matching algorithms, and analyse their matching performance in terms of different building use cases and various application-level tasks.

\section{Evaluation Plan: A Case Study in the Building Sector}

In the context of Industry 5.0 and the Internet of Things (IoT), digitisation and automation are becoming emerging research areas in the building sector. While a number of building and building-related ontologies have been developed, data interoperability issues have become more apparent. Different building ontologies are developed and maintained by different institutions. These ontologies are modelled at multiple levels of abstraction for various purposes, and their definitions are frequently competing and overlapping. Proposed ontology-compliant KGs would potentially help with the unified vision of building ontologies, where the data in this domain has complexity and variety in concepts and relations.

\section{Preliminary Results}
\label{sec: results}

\subsection{Ontology Compliance within KG}

\begin{figure}[htbp]
\centerline{\includegraphics[width=1\linewidth]{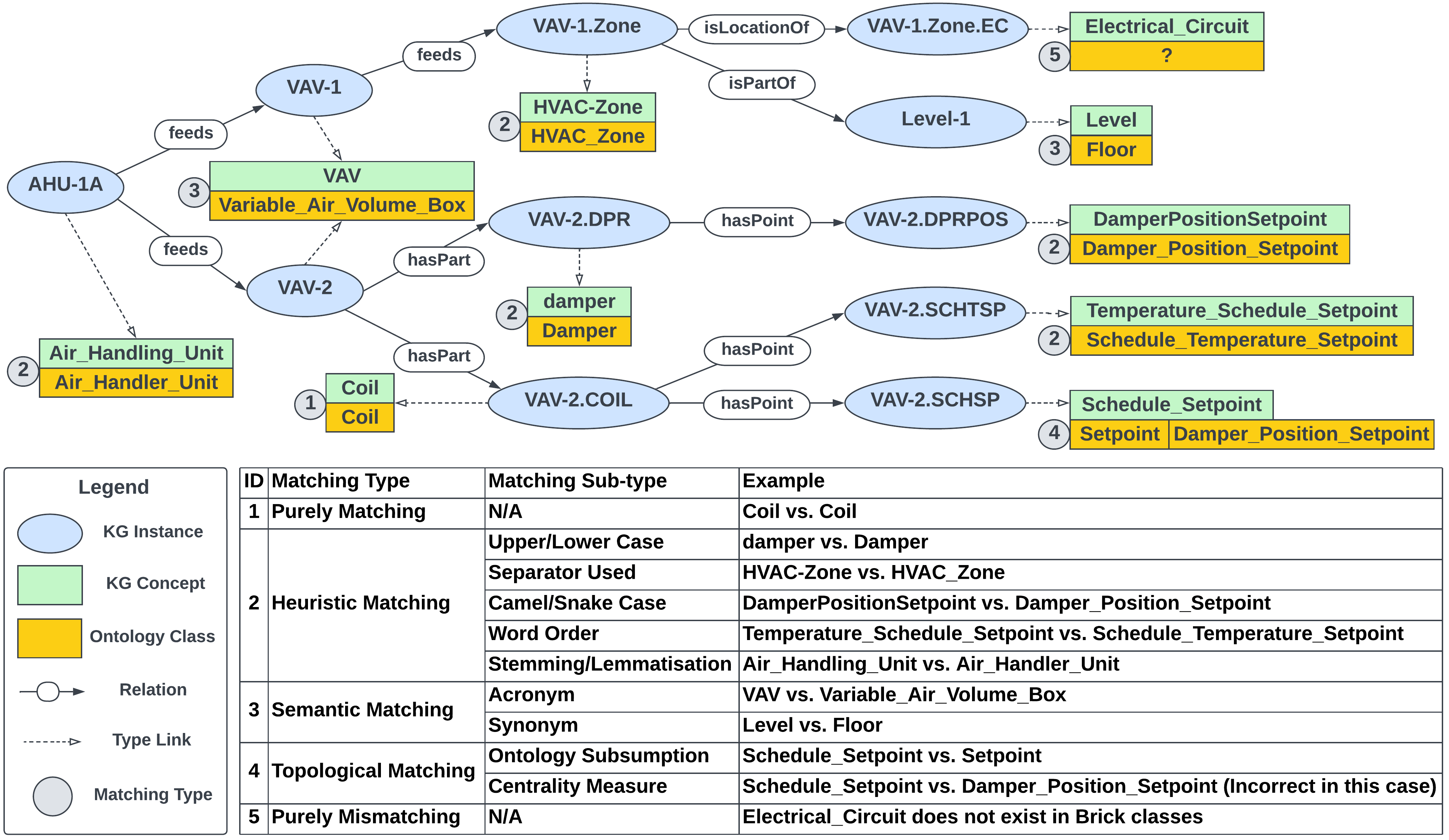}}
\caption{An example of ontology compliance within KG.}
\label{fig: within compliance}
\end{figure}

A KG and its ontology share all terms and topology. However, the concepts and properties defined in KG and ontology can be mismatched due to human errors, design choices, or changes in newer versions. Fig.~\ref{fig: within compliance} shows different types of node matching in a snippet of an air handling unit (AHU) system represented by a KG and its ontology Brick Schema~\cite{balaji2016brick} (abbr. ``Brick''). The concepts with green colours are KG classes, while the concepts with yellow colours are ontology classes. Different matching types and their examples are shown in the table below. These also applied to the property matching between KG and its ontology. We design Algorithm~\ref{alg: within compliance} for building ontology compliance within KG. It has two phases: (1) Entity Alignment and (2) Ontology Reconstruction. We first find non-compliant terms in the KG and replace them with the most relevant classes and properties in the original ontology, assigning a confidence score for each replacement. Then, we find all the related triples (including constraints and axioms) and restore the ontology hierarchies.

\begin{algorithm}[htbp]
\scriptsize

\caption{Building Ontology Compliance within KG}
\label{alg: within compliance}
\begin{multicols}{2}
\begin{algorithmic}
\renewcommand{\algorithmicrequire}{\textbf{Input:}}
\renewcommand{\algorithmicensure}{\textbf{Output:}}
\REQUIRE Ontology-based $KG$, Ontology $Onto$
\ENSURE  Reshaped Onto $Onto_{RE}$, Confidence $\mu$

\STATE \textbf{/* Phase 1: Entity Alignment */}\\
\STATE /* Find all terms in KG */
\STATE Concept Set $Con\_S$, Relation Set $Rel\_S$ $\leftarrow \varnothing$ 
\STATE $ Con\_S,Rel\_S \leftarrow findConceptAndRelation(KG) $

\STATE /* Find all classes and properties in Onto */
\STATE Class Set $Cls\_S$, Property Set $Pro\_S$ $\leftarrow \varnothing$
\STATE $ Cls\_S,Pro\_S \leftarrow findClassAndProperty(Onto) $

\STATE /* Divide each naming into a keyword set */
\STATE $ Con\_S,Rel\_S \leftarrow findKwd(Con\_S,Rel\_S) $
\STATE $ Cls\_S,Pro\_S \leftarrow findKwd(Cls\_S,Pro\_S) $

\STATE /* Match classes between KG and Onto */
\FOR {$ i \in Con\_S $}
\FOR {$ j \in Cls\_S $}
\IF {$ purelyMatching (i,j) \neq \varnothing $}
\STATE $ \mu_i \leftarrow 1 $
\ELSIF {$ heuristicMatching (i,j) \neq \varnothing $}
\STATE $ \mu_i \leftarrow Levenshtein Distance $
\STATE $ Con\_S \rightarrow i \land Con\_S \leftarrow j $
\ELSIF {$ semanticMatching(i,j) \neq \varnothing $}
\STATE $ \mu_i \leftarrow Similarity $
\STATE $ Con\_S \rightarrow i \land Con\_S \leftarrow j $
\ELSIF {$ topologicalMatching(i,j) \neq \varnothing $}
\STATE $ \mu_i \leftarrow Accuracy $
\STATE $ Con\_S \rightarrow i \land Con\_S \leftarrow j $
\ENDIF
\ENDFOR
\ENDFOR

\STATE /* Match properties between KG and Onto */\\
\FOR {$ k \in Rel\_S $}
\FOR {$ l \in Pro\_S $}
\STATE /* Corresponding procedure applies to Rel\_S and Pro\_S */
\ENDFOR
\ENDFOR

\STATE /* Calculate total confidence */
\STATE Total Confidence $\mu$ $\leftarrow \varnothing$ 
\STATE $ \mu \leftarrow Avarage (\mu_1, ..., \mu_i, \mu_1, ..., \mu_k)$

\STATE \textbf{/* Phase 2: Ontology Reconstruction */}\\
\STATE /* Find super-classes */
\STATE $ Onto\_{RE}$ Class Set $ RE\_Cls\_S \leftarrow \varnothing $
\STATE $ RE\_Cls\_S \leftarrow findSuperClasses(Con\_S) $

\STATE /* Find super-properties */
\STATE $ Onto\_{RE} $ Property Set $ RE\_Pro\_S \leftarrow \varnothing $
\STATE $ RE\_Pro\_S \leftarrow findSuperProperties(Rel\_S) $

\STATE /* Restore reshaped ontology */
\STATE Reshaped Ontology $ Onto_{RE} \leftarrow \varnothing $
\STATE $ Onto_{RE} \leftarrow findTriples(RE\_Cls\_S, RE\_Pro\_S) $

~\\

\RETURN $ Onto_{RE},\mu $
 
\end{algorithmic}
\end{multicols}
\end{algorithm}

\noindent\textbf{Evaluation} The preliminary experiment uses the sample example from the Brick Schema official website. We synthesise a number of mismatched classes and properties with different types. The results in Table~\ref{tab: within compliance} show that reshaped ontologies can significantly reduce the original ontology size and increase the number of used and matched classes. We can also observe a trade-off between the confidence score and the level of matching applied. The confidence score slightly decreases when the level of matching increases. A potential reason is that Level 3 and Level 4 use learning-based approaches. While they are more powerful at discovering more pairs of matches, the confidence level of the matching accuracy highly depends on the models and methods used.

\begin{table}[!t]
\renewcommand\arraystretch{1.2}
\tabcolsep=0.15cm
\caption{Evaluation of algorithm for building ontology compliance within KG.}
\label{tab: within compliance}
\begin{adjustbox}{width=1\linewidth,center}
\begin{tabular}{|r|c|c|c|}\hline
 Type of Ontology \& Matching Level     & Used Entity   & Matching Rate   & Confidence \\ \hline
Original Ontology \& Matching Lv. 1     & 0.52\%        & 46.15\%         & 100.00\%   \\
Reshaped Ontology \& Matching Lv. 1     & 30.00\%       & 46.15\%         & 100.00\%   \\
Reshaped Ontology \& Matching Lv. 2     & 38.46\%       & 76.92\%         & 97.50\%    \\
Reshaped Ontology \& Matching Lv. 3     & 42.31\%       & 84.62\%         & 88.00\%    \\
Reshaped Ontology \& Matching Lv. 4     & 46.15\%       & 92.31\%         & 78.33\%    \\ \hline
\end{tabular}
\end{adjustbox}
\end{table}

\subsection{Ontology Compliance over KGs} 
\begin{figure}[htbp]
\centerline{\includegraphics[width=1\linewidth]{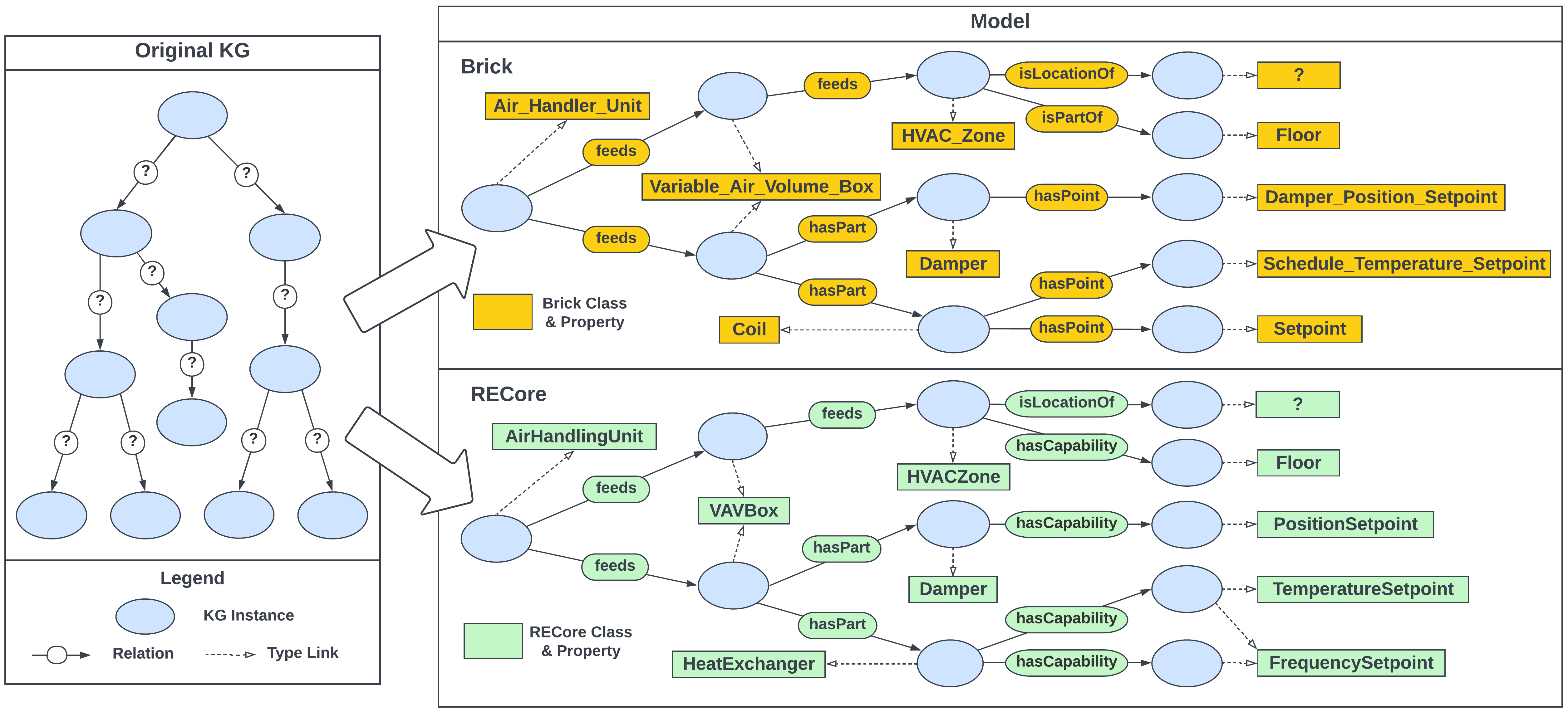}}
\caption{An example of ontology compliance over KGs.}
\label{fig: over compliance}
\end{figure}

It is often the case that a KG may be restructured to comply with one of several different ontologies, while preserving the KG's intended information content, as illustrated in Fig.~\ref{fig: over compliance}.

The key task is ontology alignment and matching. Embedding-based methods are prevalent for exploring the potential matching in graphs due to their conceptual simplicity and computational operability. However, there are several challenges when applying the embedding methods to ontology alignment and matching. Firstly, not all the classes and properties from the original ontology are useful for KG instance embedding. The unused classes and properties could be noise for matching. Secondly, the respective KGs have no connection with each other. Embedding methods are based on random sampling, meaning the vector only represents the relative position of the nodes, and the vector number sets can be different. If two graphs are not tightly connected, the embedding results are most likely to be incorrect. Thirdly, the majority of the embedding methods are targeting graphs without schema. They focus more on topological matching rather than lexicographic matching. We employ our findings from ontology-compliant within KG to address the first challenge, and use their compliance as an intermediate link to connect two graphs - the second challenge, and facilitate lexicographical order - the third challenge. We design Algorithm~\ref{alg: over compliance} for building ontology compliance over KGs. It has three phases: (1) Build ontology compliance within each KG, (2) Match terms across ontologies, and (3) Match overlapping terms. Phases 1 and 2 also reuse the Algorithm~\ref{alg: within compliance}.

\begin{algorithm}[htbp]
\scriptsize

\caption{Building Ontology Compliance over KGs}
\label{alg: over compliance}
\begin{multicols}{2}
\begin{algorithmic}
\renewcommand{\algorithmicrequire}{\textbf{Input:}}
\renewcommand{\algorithmicensure}{\textbf{Output:}}
\REQUIRE $KG_1$ and related ontology $Onto_1$, \\
$KG_2$ and related ontology $Onto_2$
\ENSURE  Matching Set $Match\_S(Onto_1,Onto_2)$,\\ Confidence Set $\mu(\mu_{within}, \mu_{over}) $

\STATE \textbf{/* Phase 1: Compliance within KG */}\\
\STATE $ Onto_{RE_1}, \mu_1 = withinComp(KG_1,Onto_1) $
\STATE $ Onto_{RE_2}, \mu_2 = withinComp(KG_2,Onto_2) $
\STATE $ \mu_{within} \leftarrow \mu_1, \mu_2 $

\STATE \textbf{/* Phase 2: Match terms across Onto */}\\
\STATE /* Create matching set */\\
\STATE Matching Set $ Match\_S(Onto_1,Onto_2) \leftarrow \varnothing $

\STATE /* Follow same procedure in Algorithm~\ref{alg: within compliance} */
\FOR {$ i \in Con\_Onto_{RE_1}, k \in Rel\_KG_1 $}
\FOR {$ j \in Con\_Onto_{RE_2}, l \in Rel\_KG_2 $}
\STATE $ Match\_S(Onto_1,Onto_2), \mu_{cmatch} \leftarrow $ Matched Con \& Rel, $ \mu_{over} \leftarrow \mu_{cmatch} $
\ENDFOR
\ENDFOR

\STATE \textbf{/* Phase 3: Match overlapping terms */}

\STATE /* Set a vector space */
\STATE Vector space $VecSpace \leftarrow \varnothing $
\STATE /* Put KGs and Onto\_{RE} */
\FOR {$ i \in KG_1, Onto_{RE_1}, KG_2, Onto_{RE_2} $}
\STATE $VecSpace \leftarrow i $
\ENDFOR

\STATE /* Put matched Onto set to build links */
\STATE $VecSpace \leftarrow Match\_S(Onto_1,Onto_2) $

\STATE /* Define embedding models */
\STATE Embedding Model $ Model \leftarrow X2Vec.train() $
\STATE /* Define vector set */
\STATE Vector Set $ Vec\_S \leftarrow Model.getEmbeddings() $

\STATE /* Add predict match for overlapping */
\STATE Predict Match $P\leftarrow \varnothing$
\FOR {Unmatched $ U \notin Match\_S(Onto_1,Onto_2) $}
\STATE $ P,\mu_{overlap} \leftarrow Vec\_S.getMostSimilar(U) $
\STATE $ Match\_S(Onto_1,Onto_2) \leftarrow (U,P) $
\STATE $ \mu_{over} \leftarrow \mu_{overlap} $
\ENDFOR

\STATE /* Summarise total confidence */
\STATE Confidence Set $\mu \leftarrow \varnothing$
\STATE $\mu \leftarrow \mu_{within}$
\STATE $\mu \leftarrow \mu_{over}$

~\\

\RETURN $ Match\_S(Onto_1,Onto_2)$, \\ 
$ \mu(\mu_{within}, \mu_{over}) $

\end{algorithmic}
\end{multicols}
\end{algorithm}

\noindent\textbf{Evaluation} The preliminary experiment is set to predict the similarity of two overlapping properties, brick:hasPoint in Brick Schema~\cite{balaji2016brick} (abbr. ``Brick'') and core:hasCapability in RealEstateCore~\cite{hammar2019realestatecore} (abbr. ``RECore''). The ground truth is that the meanings of these two properties are very similar. Their different names are due to their different views on how building points are embedded in the building. brick:hasPoint states that the building points are the measurable data points installed in the building, whereas core:hasCapability stands for the building points are the capabilities provided by the building to produce and ingest data. We employ three different vector embedding models to evaluate the top-k searches. Exp.1 uses the traditional KG embedding without ontology compliance, and Exp.2 uses our proposed compliance algorithm. Table~\ref{tab: over compliance} shows the results of the comparison in a test run. We can see Exp.2 outperforms Exp.1 in all three sample embedding models, particularly in top-1 and top-3 searches.

\begin{table}[htbp]
\renewcommand\arraystretch{1.2}
\caption{Evaluation of algorithm for building ontology compliance over KGs.}
\label{tab: over compliance}
\begin{adjustbox}{width=1\linewidth,center}
\begin{tabular}{|c|c|c|c|c|c|c|}
\hline
\multirow{2}{*}{@k} & \multicolumn{2}{c|}{DeepWalk~\cite{perozzi2014deepwalk}} & \multicolumn{2}{c|}{Node2Vec~\cite{grover2016node2vec}} & \multicolumn{2}{c|}{Struc2Vec~\cite{ribeiro2017struc2vec}} \\ \cline{2-7} 
    & Exp.1            &Exp.2                     & Exp.1            & Exp.2                     & Exp.1            & Exp.2 \\ \hline
1   & $0.03\pm0.17\%$  & $25.86\pm4.48\%$ & $0.04\pm0.20\%$  & $15.32\pm3.70\%$ & $96.88\pm1.52\%$ & $99.98\pm0.14\%$ \\
3   & $1.36\pm1.07\%$  & $57.49\pm4.86\%$ & $13.4\pm3.35\%$  & $56.65\pm5.28\%$ & $100\pm0.00\%$   & $100\pm0.00\%$   \\
5   & $7.58\pm2.57\%$  & $75.51\pm4.17\%$ & $66.54\pm4.85\%$ & $84.86\pm3.63\%$ & $100\pm0.00\%$   & $100\pm0.00\%$   \\ \hline
\end{tabular}
\end{adjustbox}
\end{table}

\subsection{Pattern-Based Compliance}

\begin{figure}[htbp]
\centerline{\includegraphics[width=1\linewidth]{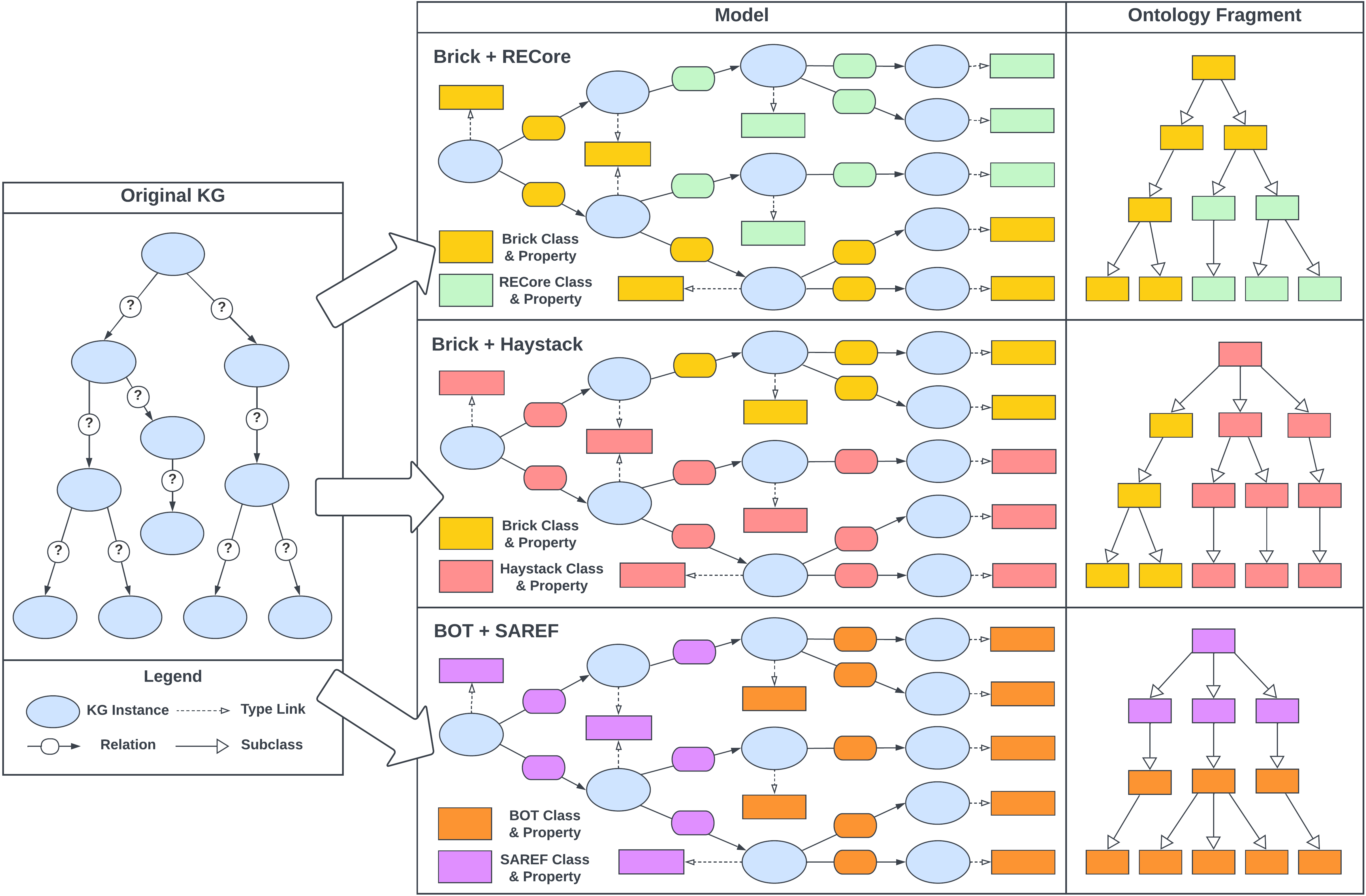}}
\caption{An example of pattern-based compliance.}
\label{fig: pattern compliance}
\end{figure}

Ontology can be decomposed into smaller ontology fragments. For example, the concepts defined in Brick Schema (abbr. ``Brick'') can be decomposed into three high-level abstraction fragments: Spaces (i.e., brick:Location), Building Equipment and Systems (i.e., brick:Equipment and brick:System), and Building Points (i.e., brick:Point). For each fragment, they can be replaced with the same concepts defined in other building ontologies, such as RealEstateCore~\cite{hammar2019realestatecore} (abbr. ``RECore'') and Project Haystack~\cite{john2020project} (abbr. ``Haystack''), or building-related domain ontologies, such as BOT~\cite{rasmussen2021bot} for spatial information, SAREF~\cite{daniele2015created} for equipment and systems, and SSN~\cite{compton2012ssn}/SOSA~\cite{haller2019modular} for building points. Fig.~\ref{fig: pattern compliance} demonstrates an example of KG represented by different combinations of building and related domain ontologies. These new ontology fragments can represent the same information as the original ontology, but they can have different numbers of classes, properties, and hierarchies. If we consider ontology fragments, the problem of ontology compliance becomes more complex.

\begin{figure*}[htbp]
\centerline{\includegraphics[width=1\linewidth]{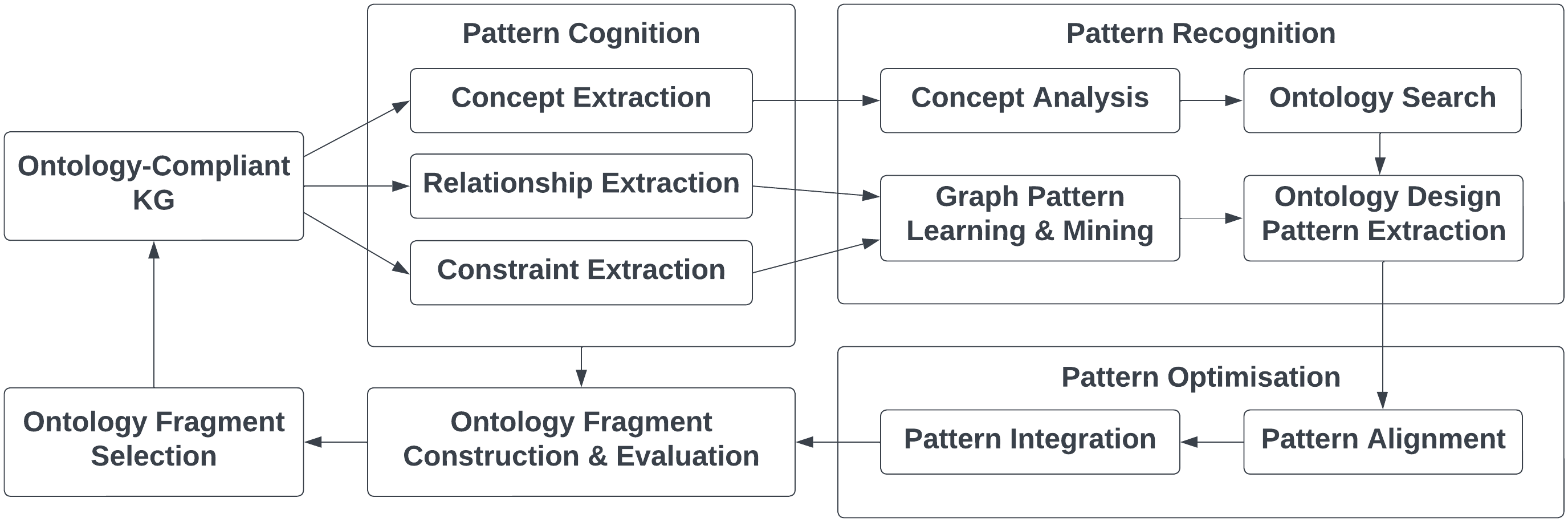}}
\caption{The architecture of building pattern-based compliance.}
\label{fig: pattern-based-architecture}
\end{figure*}

Fig.~\ref{fig: pattern-based-architecture} shows the architecture of building pattern-based compliance. It has three main components, namely (1) Pattern Cognition, (2) Pattern Recognition, and (3) Pattern Optimisation. The basic idea is to extract the concepts, relationships, and constraints from the ontology-compliant KG. Each of them goes through a learning and matching process to find their patterns. Then, we integrate and align the same or similar patterns, and use these generic patterns to reconstruct new ontology fragments.

\begin{figure}[htbp]
\centerline{\includegraphics[width=0.6\linewidth]{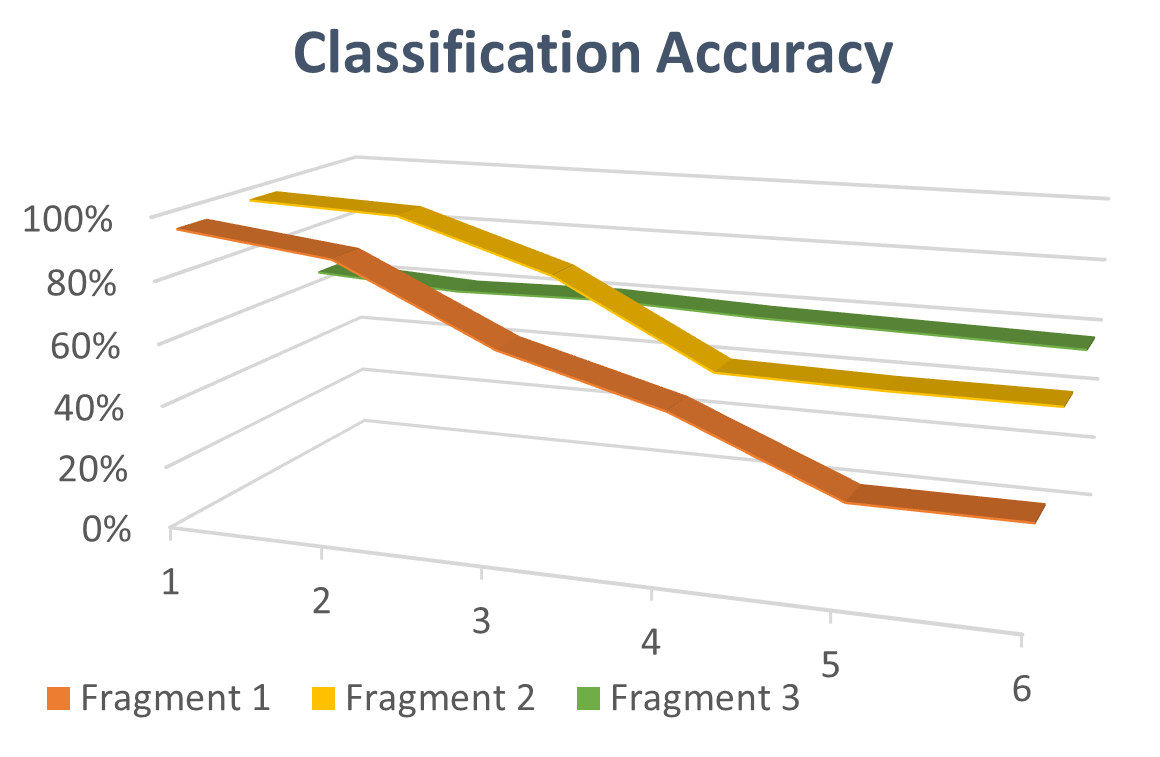}}
\caption{A joint approach using classification accuracy to evaluate the performance. The fluctuation of each ontology fragment varies at different levels of abstraction.}
\label{fig: joint approach}
\end{figure}

\noindent\textbf{Evaluation} Ontology fragments may have different numbers of namespaces, levels of abstraction, concept coverage, depth of the class hierarchy, completeness and expressiveness, and performance metrics. Liebig's law~\cite{de1994liebig} is used for ontology fragment construction, evaluation, and selection. The multi-criteria selection depends on the minimum criteria being satisfied. We also introduce a joint-learning approach to evaluate the performance of ontology fragments. An example is shown in Fig.~\ref{fig: joint approach}. Ontology Fragment 1, 2, and 3 are generated from the same KG. We fit them into the embedding model and perform the classification task according to the original KG. Based on different levels of abstraction, the classification accuracy of the KG embedding decreases at different rates. Fragment 1 and 2 have higher accuracy in Level 1 and Level 2 abstractions, but they have a significant drop in Level 3. By contrast, Fragment 3 decreases gradually at all levels of abstraction.

\noindent Results given here may have slight differences across different platforms and library versions. The code implementation is available at~\url{https://github.com/qzc438/ontology-compliant-kgs} (access will be made available on request).

\section{Conclusions}

In this paper, we present a new concept of ontology-compliant KGs, showing promising results in matching and aligning ontologies within KG and over KGs. We also illustrate our design for advanced pattern-based compliance. Further work will focus on justifying the results with the capability to allow ontology compliance on large-scale KGs, and implementing pattern-based compliance in a comprehensive framework that enables automatic ontology fragment integration, evaluation, and selection for real-world application-level KGs.

\noindent\textit{\textbf{Acknowledgements} This project is supervised by Kerry Taylor, Sergio Rodríguez Méndez, Subbu Sethuvenkatraman, Qing Wang, and Armin Haller.
The author also thanks program mentor Maria Maleshkova for providing valuable feedback.}

\bibliographystyle{splncs04}
\bibliography{qiang-bibliography-eswc}

\end{document}